\documentclass[journal=jacsat,manuscript=article]{achemso}

\usepackage[version=3]{mhchem} 
\usepackage{mathtools}
\usepackage{amsmath}
\usepackage{amssymb}
\usepackage{amsthm}
\usepackage{amsfonts}
\usepackage{braket}
\usepackage{graphicx} 

\author{Michele Guerrini}
\affiliation
{Humboldt-Universit\"at zu Berlin, Physics Department and IRIS Adlershof, 12489 Berlin, Germany}
\alsoaffiliation
{Carl von Ossietzky Universität Oldenburg, Institute of Physics, 26129 Oldenburg, Germany}
\author{Enrique Delgado Aznar}
\affiliation{Humboldt-Universit\"at zu Berlin, Physics Department and IRIS Adlershof, 12489 Berlin, Germany}
\author{Caterina Cocchi}
\affiliation
{Humboldt-Universit\"at zu Berlin, Physics Department and IRIS Adlershof, 12489 Berlin, Germany}
\alsoaffiliation
{Carl von Ossietzky Universität Oldenburg, Institute of Physics, 26129 Oldenburg, Germany}
\email{caterina.cocchi@uni-oldenburg.de}

\title{Electronic and Optical Properties of Protonated Triazine Derivatives}

\begin{document}

\newpage

\begin{abstract}
The peculiar electronic and optical properties of covalent organic frameworks (COFs) are largely determined by protonation, a ubiquitous phenomenon in the solution environment in which they are synthesized.
The resulting effects are non-trivial and appear to be crucial for the intriguing functionalities of these materials.
In the quantum-mechanical framework of time-dependent density-functional theory, we investigate from first principles the impact of protonation of triazine and amino groups in molecular building blocks of COFs in water solution.
In all considered cases, we find that proton uptake leads to a gap reduction and to a reorganization of the electronic structure, driven by the presence of the proton and by the electrostatic attraction between the positively charged protonated species and the negative counterion in its vicinity. 
Structural distortions induced by protonation are found to play only a minor role.
The interplay between band-gap renormalization and exciton binding strength determines the energy of the absorption onsets: when the former prevails on the latter, a red-shift is observed. 
Furthermore, the spatial and energetic rearrangement of the molecular orbitals upon protonation induces a splitting of the lowest-energy peaks and a decrease of their oscillator strength in comparison with the pristine counterparts. 
Our results offer quantitative and microscopic insight into the role of protonation on the electronic and optical properties of triazine derivatives as building blocks of COFs. As such, they contribute to rationalize the relationships between structure, property, and functionality of these materials.
\end{abstract}

\newpage
\section{Introduction}
Covalent organic frameworks (COFs) are a novel class of porous materials formed by organic building blocks connected together by covalent bonds~\cite{feng+12csr}. 
These systems, synthesized so far both in the two- and three-dimensional form~\cite{cote+05sci,elka+07sci,jin+17sci,Kim+2,noda+19acie}, have been regarded with particular interest in the last few years, due to their potential for a number of applications~\cite{ding+13csr}, ranging from gas storage~\cite{li+14cc,wu+17chpl,zhu+17,kona+18cgd} to catalysis~\cite{xu+14cc,lin+15sci,wang+16jacs}, and from chemical sensing~\cite{das+15cs,gao+18cc,BojdysCOF,meng+19jacs,Xu+6njc2018} to optoelectronics~\cite{chen+14jacs,cai+14cs,xie+15jmcc,huan+15acie}.
One of the key properties of COFs is their chemical versatility~\cite{huan+16natrm}: Even slight modifications in the composition of the backbone structure can lead to substantial variations of the intrinsic properties of the materials and, consequently, to a modulation of their functionalities~\cite{spit+12acie,jin+15jacs,xu+16cc,zeng+16am}. 

Proton uptake typically influences the pH of the network and enhances, for example, its ability to adsorb gaseous molecules~\cite{liu+12jcis,babu+19acsami,chen+19tac}.
Some COFs even exhibit switching capabilities driven by the protonation of specific nitrogen-containing chemical units, such as triazine and amino groups~\cite{liu+12jcis,byun+16natcom,rao+17jacs,BojdysCOF}: Protonation of triazine groups was recently identified as the driving mechanism for the peculiar optical activity of layered COFs formed by 1,3,5-tris-(4-aminophenyl) triazine (TAPT) building blocks linked through the keto-enamine bridge-units (1,4-phenylene)bis(3-hydroxyprop-2-en-1-one) (PBHP)~\cite{BojdysCOF}. 
The interaction with \ce{HCl}, in gas phase or in solution, induces a red-shift of both optical absorption and emission. Interestingly, these processes can be fully reversed through the introduction of \ce{NH3}, which deprotonates the system and restores its original electronic and optical response. This chemical cycle can be exploited to design novel chemical sensors~\cite{wang+19analc,asch+18natcom,asch+19jacs,Zhang+6cc2020,asch+19jacs}. 

The experimental research carried out on (protonated) COFs as well as on their properties and functionalities has not yet been corresponded by robust theoretical analysis that is able to fully rationalize the underlying physical mechanisms. 
Atomistic studies on these materials have been primarily devoted to identify stable structures~\cite{luko+11cej,gont+17acsami,shar+18iecr} and to model processes related to gas uptake~\cite{han+09csr,guan+19cej}.
From the chemistry side, several \textit{ab initio} studies have been performed to complement experimental analysis of protonation in N-heterocyclic molecules~\cite{desp+07njc,herr+14joc,racz+17jpca,bari19acsomega,mana+19cpl}. 
This body of work, however, was mainly focused on the description and rationalization of acid-base reactions~\cite{racz+16cr} rather than on the understanding of the effects of protonation on the opto-electronic activities of these compounds.
To date, very limited efforts have been dedicated to theoretical investigations on the peculiar optical response of COFs and on the effects induced by protonation. 
The complexity of the involved processes and the need for advanced quantum-mechanical descriptions to capture them have certainly played a role in hindering this progress.

Herein, we present a first-principles study that rationalizes the electronic and optical properties of triazine derivatives often employed as building blocks in COFs.
We adopt the quantum-mechanical framework of time-dependent density-functional theory to gain a fundamental understanding on how the electronic structure of these systems and their ability to absorb light are affected by proton uptake. 
To this end, we focus on the TAPT molecule protonated by dissociated \ce{HCl} in solution and examine the effects of protonation on triazine and amino groups.
For comparison, we consider two additional molecules, namely 1,3,5-tris(4-aminophenyl)benzene (TAP) and 2,4,6-triphenyl-1,3,5-triazine (TPT), which have a very similar structure as the TAPT but include either the amino (TAP) or the triazine (TPT) group only.
We investigate the impact of protonation on the electronic structure of these molecules, focusing on the effects induced on the electronic gap and on the frontier states.
We analyze the optical properties in terms of absorption spectra and excitation energies, as well as of structural distortions driven by the attached proton.

\section{Methodology}
\subsection{Theoretical Background}

The results of this work are obtained from time-dependent density functional theory (TDDFT) in the linear-response scheme~\cite{casi96} coupled to the polarizable continuum model (PCM)~\cite{camm-menn99jcp} in order to take into account solvent effects implicitly.
In this framework, the molecule, treated atomistically, is enclosed in a cavity surrounded by a continuous medium solely defined by its dielectric function in the static limit, $\varepsilon_0$, as well as in the high-frequency limit, $\varepsilon_\infty$.

As a starting point, the ground-state properties of the solute molecule are obtained from the time-independent Kohn-Sham (KS) equations~\cite{kohn-sham65pr}, which, coupled to the PCM and in atomic units, read:
\begin{equation}
\left[-\frac{1}{2}\nabla^2 + V_{Hxc}[\rho](\textbf{r}) + V_{PCM}[\rho](\textbf{r}) +V_{ext}(\textbf{r})\right]\phi_i^{KS}(\textbf{r})=\epsilon_i\phi_i^{KS}(\textbf{r}).
\label{KSeq}
\end{equation}
In Eq.~\eqref{KSeq}, $\rho(\textbf{r})=\sum_{i=1}^N|\phi_i^{KS}(\textbf{r})|^2$ is the electronic density of the $N$-electron system, with $\phi_i^{KS}(\textbf{r})$ and $\epsilon_i$ being the $i^{th}$ KS eigenfunction and eigenenergy, respectively;  $V_{Hxc}[\rho](\textbf{r})$ is the Hartree exchange-correlation (Hxc) potential which takes into account electron-electron interactions; $V_{ext}(\textbf{r})$ is the external potential exerted by the nuclei to the electrons. 
The term $V_{PCM}[\rho](\textbf{r})$, which is also implicitly a functional of the electronic density, is the effective potential describing the interaction between solute and solvent, with the latter modelled as a continuum medium.\cite{canc+97jcp,camm-menn99jcp,delg+15jcp}

In order to access excited state properties, one needs to extend Eq.~\eqref{KSeq} to the time domain~\cite{rung-gros84prl}.
In the adopted linear response TDDFT formalism~\cite{casi96}, the following generalized eigenvalue problem has to be solved:
\begin{equation}
\begin{pmatrix} \textbf{A}&\textbf{B} \\ \textbf{B}^*&\textbf{A}^*  \end{pmatrix}
\begin{pmatrix} \textbf{X} \\ \textbf{Y}  \end{pmatrix}
=\omega_i
\begin{pmatrix} \textbf{1}&\textbf{0} \\ \textbf{0}&\textbf{-1}  \end{pmatrix}
\begin{pmatrix} \textbf{X}\\ \textbf{Y} \end{pmatrix},
\label{casida}
\end{equation}
where the eigenvalues $\omega_i$ correspond to the energy of the $i^{th}$ excited state, and the eigenvectors $X^i_{ak}$ and $Y^i_{ak}$ represent the weights associated to the resonant and anti-resonant (single-particle) transitions between the occupied state $k$ and the unoccupied state $a$, respectively.
In Eq.~(\ref{casida}),
\begin{equation}
A_{ka,jb}=\delta_{a,b}\delta_{k,j}(\epsilon_a-\epsilon_k) + \int\phi^*_k(\textbf{r})\phi_a(\textbf{r})[K_{Hxc}(\textbf{r,r'})+K_{PCM}(\textbf{r,r'})]\phi_b^*(\textbf{r'})\phi_j(\textbf{r'})d^3rd^3r', 
\label{aker}
\end{equation}
and
\begin{equation}
B_{ka,jb}=A_{ka,bj},
\label{bker}
\end{equation}
where the indexes $k$ and $j$ ($a$ and $b$) run over occupied (unoccupied) states, and where $K_{Hxc}(\textbf{r,r'})=\frac{\delta V_{Hxc} (\textbf{r'})}{\delta \rho (\textbf{r})}$ and $K_{PCM}(\textbf{r,r'})=\frac{\delta V_{PCM} (\textbf{r'})}{\delta \rho (\textbf{r})}$ are the Hxc and the PCM interaction kernels, respectively. 
Eigensolutions of Eq.~\eqref{casida} yield the oscillator strengths $S_i=\frac{2}{3}\omega_i||\sum_{ak}C_{ak}^{i}\textbf{d}_{ak}(\textbf{r}) + D^i_{ka}\textbf{d}^*_{ak}(\textbf{r})||^2$, where $\textbf{d}_{ak}=\int\phi^{*}_a(\textbf{r})\,\textbf{r}\,\phi_k(\textbf{r})$ is the transition dipole moment between the occupied KS state (superscript omitted) $k$ and the unoccupied one $a$.
The optical absorption is calculated as
\begin{equation}
    Abs(\omega)=\sum_{i}S_iF(\omega - \omega_i),
    \label{os}
\end{equation}
where $F(\omega)$ is the broadening function. 

\subsection{Computational Details}
All calculations are performed with the software Gaussian16~\cite{g16} in the framework of spin-restricted (TD)DFT with enforced singlet spin multiplicity, using the cc-pVTZ basis set and the Grimme-D3 scheme~\cite{GrimmeD3} to include dispersion interactions. The range-separated hybrid functional CAM-B3LYP~\cite{YANAI200451}, which is well established in the studies of light-harvesting molecules and complexes~\cite{koba-amos06cpl,zhu+11cm,pedo13jctc,blas+18csr,vale-cocc19jpcc,vale+20pccp,schi+20jpcc,arvi+20jpcb}, is employed to approximate the exchange-correlation potential. Due to its high fraction of exact exchange, this functional additionally alleviates the self-interaction error, which is known to be particularly problematic for charged systems~\cite{ruzs+07jcp,vydr+07jcp,korz+08jcp,schm-kuem16prb}. 
All calculations are performed at zero temperature, hence, entropic effects are not included.

Both pristine and protonated molecules are considered in water solvent, modeled within the integral-equation PCM formalism~\cite{canc+97jcp} by the static and high-frequency dielectric constants $\varepsilon_0=78.36$ and $\varepsilon_\infty=1.78$, respectively. 
The solute molecule is embedded in a cavity with shape defined by interlocking spheres centered at each atomic position~\cite{delg+15jcp}. 
Solvation energies are explicitly taken into account in both single-point calculations, including the evaluation of formation energies, and in the structural optimizations.
The geometries of the neutral species are relaxed until the residual interatomic forces are smaller than $10^{-5}$~Ry/bohr without imposing any symmetry constraint. 

Protonated species are modeled through the interaction between their neutral counterparts and a dissociated HCl molecule in acqueous solution.
Being a strong acid, HCl in water very likely splits into H$^+$ and Cl$^-$ ions.
Hence, protons are available to be attached to the N-groups of the COF units, with the abundant Cl$^-$ anions being attracted to the positively charged protonated species.
Based on these premises, we consider here protonated molecules interacting with a Cl$^-$ counterion in their proximity, which further ensures the charge neutrality of the whole system. The position of the Cl$^-$ counterion, relative to the protonated molecule, is obtained with a structural optimization of the whole complex carried out using DFT.
For comparison, the electronic structure of the positively charge protonated species is also examined in the absence of Cl$^-$ (see details in the Supporting Information).

Chemisorption energies upon proton uptake are computed as the difference between the total energy of the protonated species in the presence of Cl$^-$ and the total energies of the isolated neutral species: $E_{chem} = E_{tot}^{mol:H^+ + Cl^-} - E_{tot}^{mol} - E_{tot}^{HCl}$ (with $mol$ = TAPT, TPT, TAP).
The Mulliken charge analysis~\cite{mull55jcp} is used to evaluate the charge redistribution between the pristine and protonated molecules.

The accuracy of the employed hybrid TDDFT formalism to describe the optical excitations is benchmarked for the pristine molecules \textit{in vacuo} against the corresponding results obtained from many-body perturbation theory~\cite{brun12jcp,brun+13jctc} ($GW$ approximation and Bethe-Salpeter equation) using the MOLGW code.~\cite{molgw} The good agreement between the optical spectra computed with the two approaches is reported in the Supporting Information.

\begin{figure}
	\centering
	\includegraphics[width=1.0\textwidth]{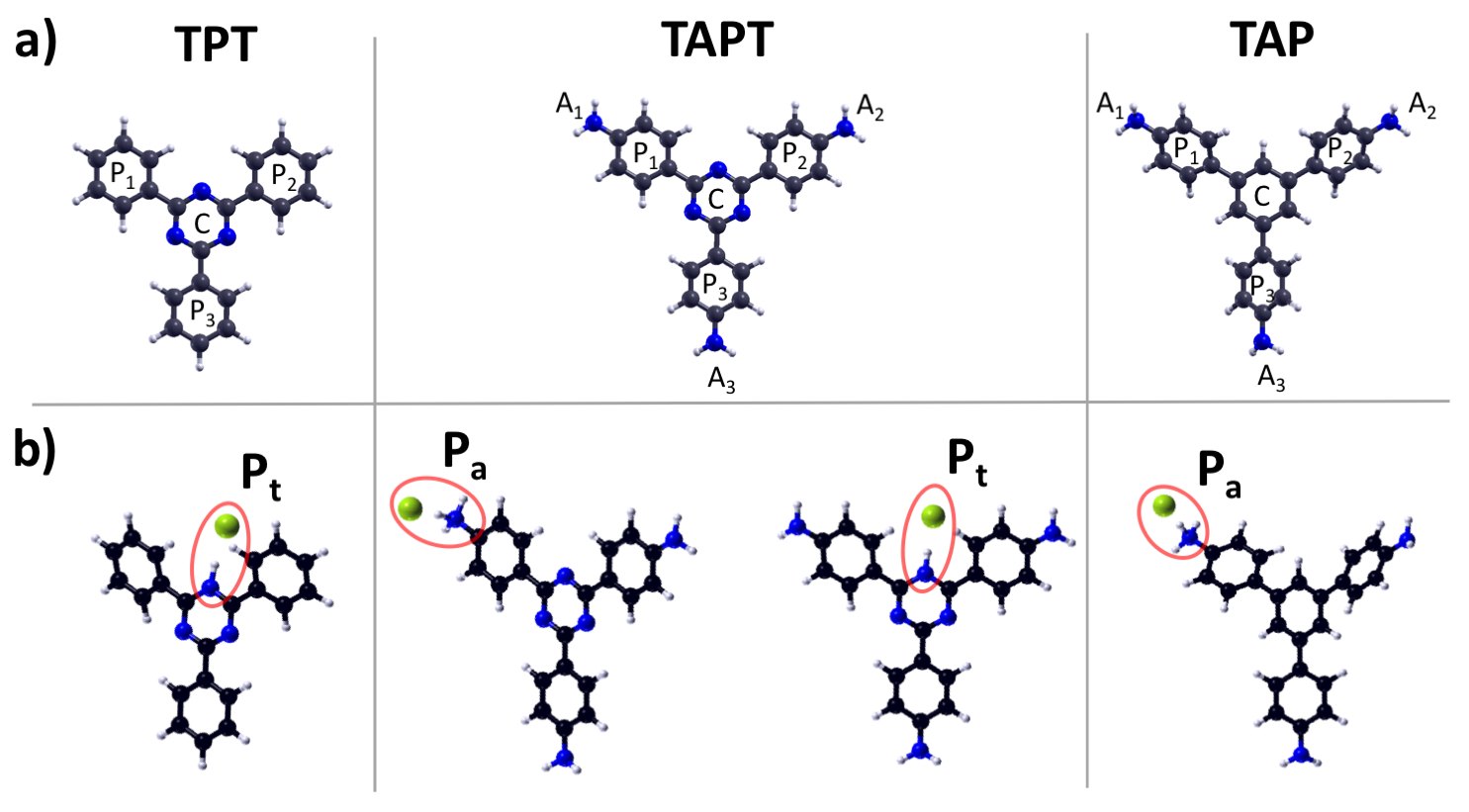}
	\caption{(a) Ball-and-stick representations of the equilibrium geometries of TPT, TAPT, and TAP. In each molecule, phenyl rings (P$_{1-3}$), amino groups (A$_{1-3}$), and the core ring (C) are marked. (b) Protonated species with the site of proton uptake (triazine ring, $P_t$, or amino group, $P_a$) and the residual chlorine anion circled in red. Carbon, nitrogen, hydrogen, and chlorine atoms are depicted in black, blue, white, and green, respectively.}
	\label{systems}
\end{figure}

\section{Results}

\subsection{Structures, Energetics, and Charge Distribution}

The three COF building blocks considered in this work, namely TAPT, TPT, and TAP (see Figure~\ref{systems}a), are characterized by a carbon-based backbone including nitrogen atoms that crucially influence their geometry and electronic structure. 
The TPT, formed by a triazine core bonded to three phenyl rings, exhibits a planar structure belonging to the $C_{3v}$ point-symmetry group.
The TAPT, which includes a central triazine ring in addition to amino-funcitonalized phenyl peripheries, also exhibits a planar backbone, due the strong CH $\cdots$ N intramolecular interactions between triazine and phenyl rings~\cite{jack+13jacs}.
On the other hand, the TAP, which comprises a H-terminated carbon-based core in the center and amino groups at the edges, is characterized by a distorted backbone with a dihedral angle of 37$^{\circ}$, due to the steric hindrance between the H atoms terminating the carbon rings in the center and in the periphery of the molecule.

The triazine- and amino-functionalized molecules considered herein behave as Lewis bases and, when exposed to dissociated \ce{HCl} in water solution, they are protonated. 
In the case of TAPT, two protonation sites are considered, corresponding to the triazine ring ($P_t$) and to one of the amino groups ($P_a$, see Figure~\ref{systems}b).
Evidently, only one protonation site is accessible in TPT ($P_t$) and TAP ($P_a$). 
Proton uptake at $P_t$ induces a backbone distortion in both TPT and TAPT, which assume dihedral angles of 27$^{\circ}$ and 20$^{\circ}$, respectively.
On the other hand, the proton uptake at $P_a$ does not induce any substantial structural deviation neither in the TAPT nor in the TAP, compared to the pristine geometries shown in Figure~\ref{systems}a).

Protonation in water is an energetically favorable process in all examined cases. 
However, the corresponding chemisorption energies differ depending on the protonation site. 
In the TAPT, protonation on $P_t$ is more favorable than on $P_a$, with $E_{chem}^{P_t-TAPT}=-925$~meV and $E_{chem}^{P_a-TAPT}=-680$~meV, respectively.
These values can be compared with those obtained in TPT and TAP protonated on the only available sites, namely $P_t$ and $P_a$, respectively: $E_{chem}^{P_t-TPT}=-626$~meV and $E_{chem}^{P_a-TAP}=-190$~meV. 
These results imply two important consequences: First, the protonation of the triazine ring is generally more favorable than the protonation of an amino group.
Second, the presence of the triazine core contributes to stabilizing the protonated system, even though it is not directly involved in the process of $H^+$ uptake. 
Both effects are caused by the electron-withdrawing character of the triazine group~\cite{elbe+00jpca,garv+11jasms,huo+19acsomega}.

\begin{table*}
\centering
\caption{Mulliken partial charges, in units of electrons, where positive (negative) values indicate electron depletion (accumulation), within the indicated molecular regions (see Figure~\ref{systems}) of pristine and protonated TAPT, TPT, and TAP in water solution. } 
\label{mulliken}%
\begin{center}
\begin{tabular}{c|c||c|c|c||c|c|c||c|c}
& \textbf{C} & \textbf{P$_1$} & \textbf{P$_2$} & \textbf{P$_3$} & \textbf{A$_1$} & \textbf{A$_2$} & \textbf{A$_3$} & \textbf{H$^+$} & \textbf{Cl$^-$} \\\hline\hline
$TAPT$  & -0.66 & +0.16 & +0.16 & +0.16 & +0.06 & +0.06 & +0.06 & $-$ & $-$  \\\hline 
$P_t-TAPT$ & -0.32 & +0.22 & +0.22 & +0.21 & +0.10 & +0.10 & +0.10 & +0.22 & -0.85 \\\hline 
$P_a-TAPT$ & -0.59 & +0.32 & +0.17 & +0.17 & +0.41 & +0.07 & +0.06 & +0.21 & -0.82  \\\hline \hline
$TPT$ & -0.48  & +0.16 & +0.16 & +0.16 & $-$ & $-$ & $-$ & $-$ & $-$  \\\hline 
$P_t-TPT$ & -0.11 & +0.24 & +0.24 & +0.22 & $-$ & $-$ & $-$ & +0.23 & -0.82 \\\hline \hline
$TAP$ & -0.06 & -0.08 & -0.08 & -0.08 & +0.10 & +0.10 & +0.10 & $-$ & $-$ \\\hline 
$P_a-TAP$ & -0.09 & +0.20 & +0.02 & +0.02 & +0.40 & +0.03 & +0.03 & +0.21 & -0.82 \\\hline \hline
\end{tabular}
\end{center}
\end{table*}

To gain further understanding on the influence of  protonation on the electronic distribution within the molecular COF units, we analyze the Mulliken partial charges as reported in Table~\ref{mulliken}. 
The results obtained for the TAPT and TPT neutral species reveal the presence of intramolecular charge transfer triggered by the pronounced electron affinity of the triazine ring~\cite{elbe+00jpca,garv+11jasms,huo+19acsomega}, which withdraws a considerable fraction of $\pi$-electron density (-0.66~$e$ in TAPT and -0.48~$e$ in TPT) from the neighboring groups.
The presence of three additional electron-donating amino groups in TAPT leads to the accumulation of extra -0.18~$e$ within the triazine ring, which explains its more negative charge compared to TPT.
The electron-donating nature of the terminating \ce{-NH2} groups gives rise to a (slight) charge unbalance also in the TAP. 
The positive charges (+0.10~$e$) in the amino groups are counterbalanced by negative charges in all the other portions of the molecule, including the central carbon ring.

The charge distribution described above is strongly influenced by protonation.
In all cases summarized in Table~\ref{mulliken}, a very large fraction of negative charge (-0.80~$e$) is localized on the chlorine counterion, as expected. 
Conversely, only about +0.20~$e$ remains on H$^+$, due to the overall charge redistribution within the protonated COF units.
Significant differences are associated to the protonation site. 
In both TAPT and TPT, the proton uptake at $P_t$ leads to a substantial reduction of the negative charge on the triazine core ring and to an enhancement of the positive charge on the neighboring functional groups.
Interestingly, the presence of the electron-donating amino groups in the TAPT does not alter the amount of positive partial charge in the phenyl rings, but limits the reduction of the negative charge in the triazine core compared to the TPT.

Protonation on the $P_a$ site drastically increments the positive charge on the involved amino group up to 0.40~$e$, and also on the phenyl connected to it. 
In the TAP, all phenyl rings assume positive partial charges upon protonation, such that the slightly negative charge on the central carbon ring is almost unaffected.
In the TAPT, the negative charge on the triazine unit is only marginally reduced to -0.59~$e$ as a compensation of the extra positive charge accumulated on the amino-functionalized phenyl ring.

\subsection{Electronic Properties}
\begin{figure}
	\centering
	\includegraphics[width=0.5\textwidth]{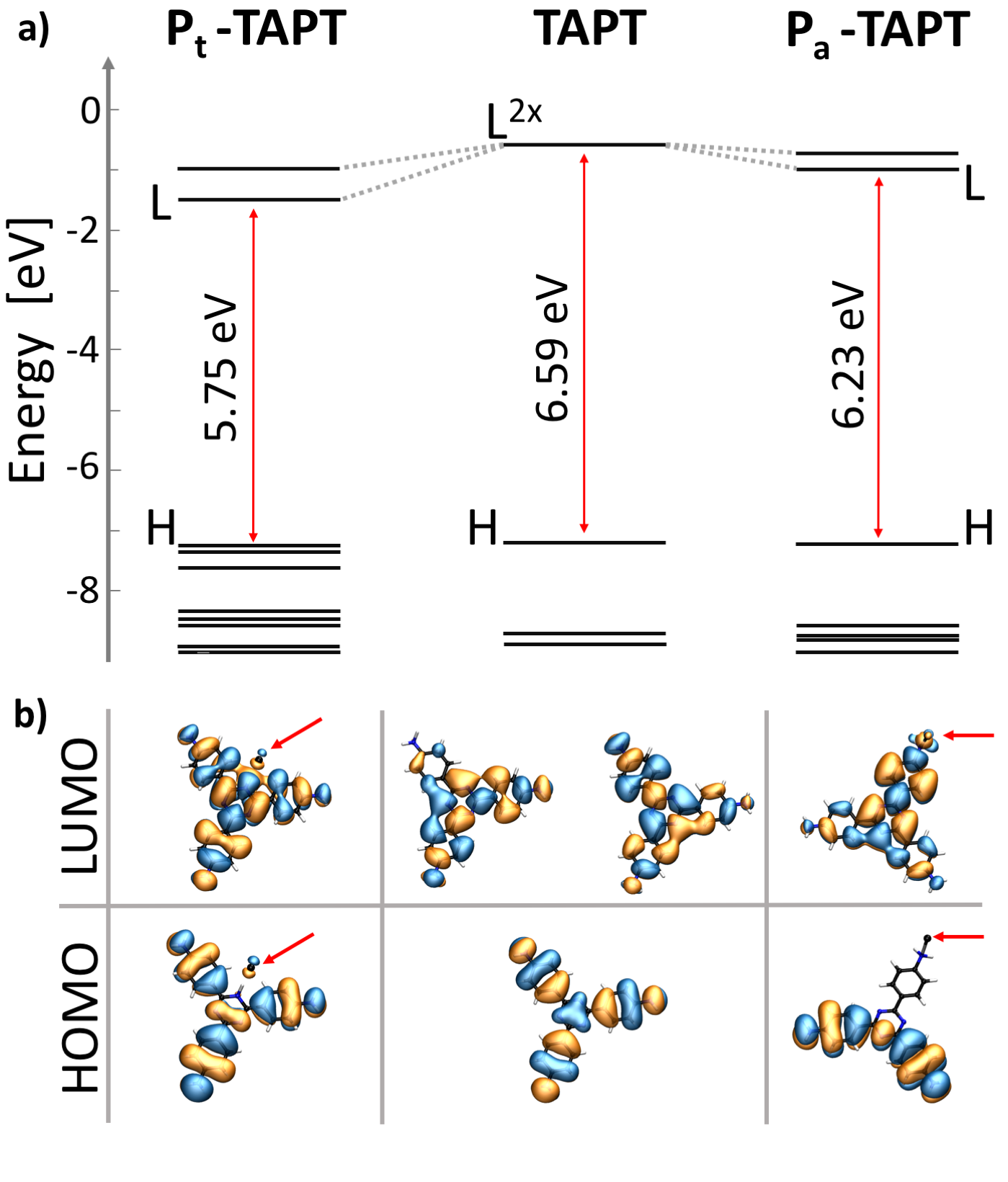}
	\caption{a) Energy levels close to the frontier of pristine (center) and protonated TAPT on the $P_t$ (left) and $P_a$ site (right) in water. The gap, given by the energy difference between the LUMO (L) and the HOMO (H), is marked by a red arrow, and the corresponding value is reported. In the pristine TAPT, the double degeneracy of the LUMO is indicated. (b) Isosurfaces of the HOMO and the LUMO with isovalue fixed at 0.01~bohr$^{-3}$. Red arrows indicate the position of the Cl$^-$ anion in proximity of the protonated COF units.}
	\label{elestr}
\end{figure}

The electronic structure of the COF units is also significantly affected by protonation.
The pristine TAPT is characterized by a double-degenerate lowest-unoccupied molecular orbital (LUMO) separated from the highest-occupied one (HOMO) by a gap of 6.59~eV (see Figure~\ref{elestr}a).
Upon proton uptake, the degeneracy of the LUMO is lifted and the electronic gap decreases regardless of the protonation site (see Table~\ref{gaps}). The gap reduction is more pronounced when the H$^+$ is attached to the triazine ring ($\Delta E_{gap}=-$0.84~eV) compared to the amino group ($\Delta E_{gap}=-$0.36~eV). 
Interestingly, this reduction is almost entirely due to the lowering of the LUMO, with the energy of the HOMO remaining almost unchanged.
On the other hand, the occupied levels below the HOMO in the $P_t$-protonated TAPT are more significantly affected by protonation than in the $P_a$-protonated TAPT, due to the prominent distorsion of the molecular backbone therein (see Figure \ref{elestr}a).

\begin{figure}
	\centering
	\includegraphics[width=0.5\textwidth]{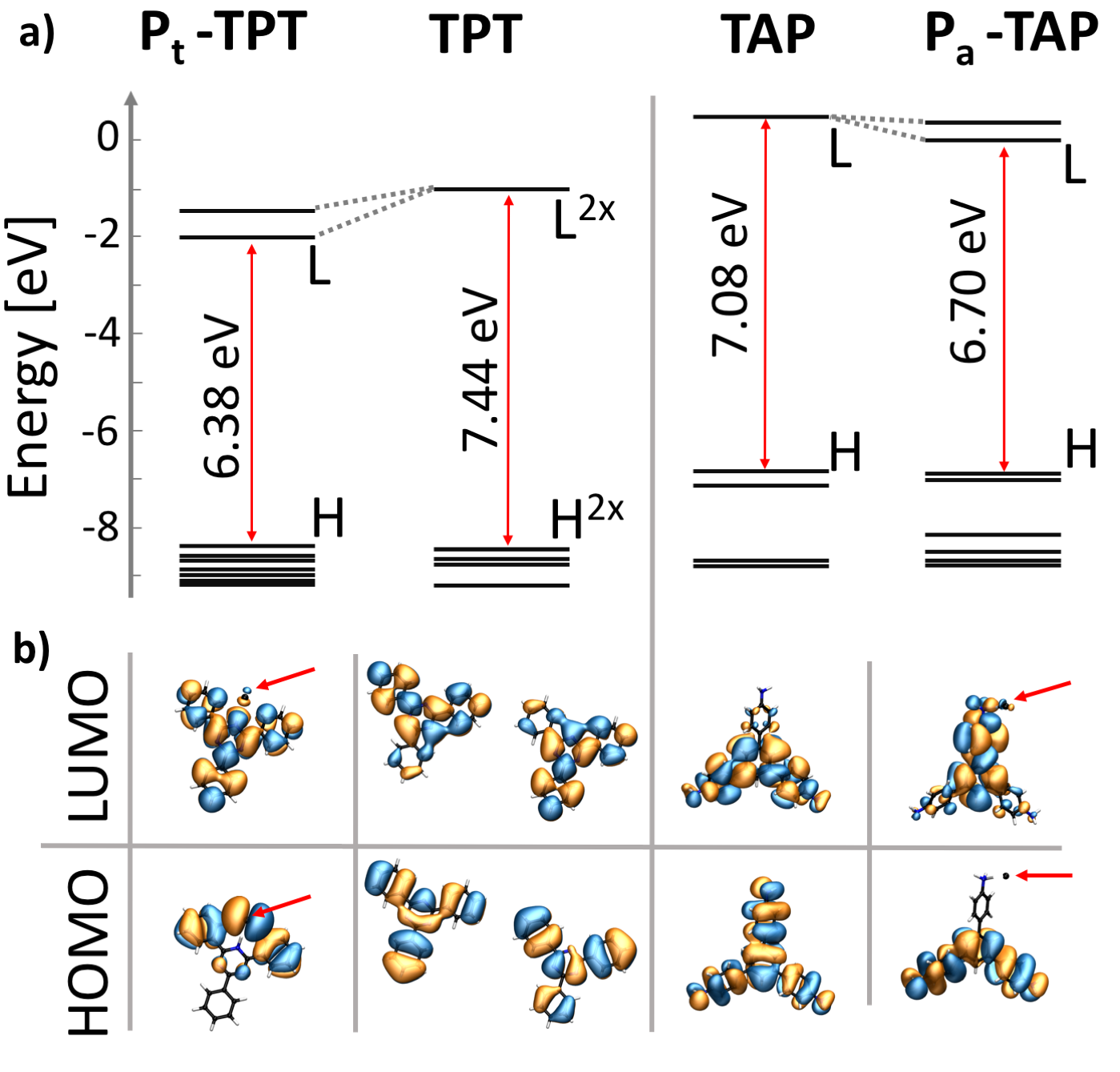}
	\caption{a) Energy levels close to the frontier of pristine and protonated TPT (left) and TAP (right) in water. The gap, given by the energy difference between the LUMO (L) and the HOMO (H), is marked by a red arrow, and the corresponding value is reported. In the pristine TPT, the double degeneracy of the HOMO and the LUMO is indicated. (b) Isosurfaces of the HOMO and the LUMO with isovalue fixed at 0.01~bohr$^{-3}$. Red arrows indicate the position of the Cl$^-$ anion in proximity of the protonated COF units.}
	\label{elestr2}
\end{figure}

The pristine TPT (see Figure~\ref{elestr2}a) is characterized by double degenerate HOMO (-8.46~eV) and LUMO (-1.02~eV).
Compared to the TAPT, the gap is about 0.80~eV larger and the frontier orbitals are down-shifted (this effect is significantly more pronounced in the HOMO).
These behaviors are ascribed to the absence of the electron-donating amino groups, which are known to increase the energies of the frontier states~\cite{cocc+11jpcl,cocc+12jpcc}.
Protonation of the TPT breaks the degeneracy of both HOMO and LUMO, and leads to a gap reduction of 1.06~eV compared to the pristine species, mainly due to the lowered LUMO energy (see Table~\ref{gaps}).  

The pristine TAP has a larger electronic gap than the TAPT.
However, the frontier orbitals of the TAP are up-shifted compared to the TAPT, due to the absence of the electron-withdrawing triazine ring. 
Protonation of an amino group leads to a gap reduction of 0.38~eV with respect to the pristine TAP.
Interestingly, this is almost the same amount of energy as in the $P_a$-protonated TAPT (see Table~\ref{gaps}).

It is evident from this analysis that protonation via dissociated HCl in water brings about a systematic reduction of the electronic gap, which is more pronounced when the proton uptake takes place on the triazine core. 
This behavior is driven by two concomitant mechanisms occurring between the protonated molecule and the chlorine counterion.
First, dissociation of HCl into H$^+$ and Cl$^-$ ions and the consequent proton uptake by the COF units leads to the formation of a dipole moment (and also of the higher order multipole moments) in the latter, which affects the electrostatics of the whole system.
The corresponding values, ranging between 12 and 14~Debye, are reported in the Supporting Information, Table~S3.
Second, the frontier orbitals of the protonated COF units interact (\textit{i.e.}, hybridize) with the 3$p$ atomic orbital of Cl$^-$ (see Figure~\ref{elestr}b and Figure~\ref{elestr2}b), which is found at -8.02~eV. 
Except for the HOMO of $P_a$-protonated TAPT and TAP, in all the occupied frontier orbitals of the protonated molecules the electron density is also delocalized on the chlorine anion. 

\begin{table*}
\centering
\caption{Absolute values and relative variations with respect to the pristine species of the Kohn-Sham electronic gaps, and Kohn-Sham energies of the frontier orbitals (both in eV) of the neutral and protonated COF units, including the Cl$^-$ counterion. The prefix $D_a$- and $D_t$- indicate neutral systems in the (distorted) geometries obtained upon protonation. } 
\label{gaps}%
\begin{center}
\begin{tabular}{c||c|c||c|c||c|c}
\textbf{Species} & $E_g$ & $\Delta E_g$ & HOMO & LUMO & $\Delta$HOMO & $\Delta$LUMO  \\\hline\hline
$TAPT$  & 6.59 & $-$ & -7.17 & -0.58 & $-$ & $-$ \\ \hline
$P_t-TAPT$ & 5.75 & -0.84 & -7.24 & -1.49 & -0.07 & -0.91  \\
$D_t-TAPT$ & 6.43 & -0.16 & -7.03 & -0.60 & 0.14 & -0.02 \\\hline 
$P_a-TAPT$ & 6.23 & -0.36 & -7.20 & -0.97 & -0.03 & -0.39  \\
$D_a-TAPT$ & 6.54 & -0.05 & -7.17 & -0.63 & 0.00 & -0.05 \\\hline \hline
$TPT$ & 7.44 & $-$ & -8.46 & -1.02 & $-$ & $-$ \\ \hline
$P_t-TPT$ & 6.38 & -1.06 & -8.38 & -2.00 & +0.08 & -0.98 \\
$D_t-TPT$ & 7.35 & -0.09 & -8.43 & -1.07 & +0.03 & -0.05\\\hline \hline
$TAP$  & 7.08 & $-$ & -6.67 & 0.41 & $-$ & $-$ \\ \hline
$P_a-TAP$ & 6.70 & -0.38 & -6.75 & -0.05 & -0.08 & -0.46 \\
$D_a-TAP$ & 7.11 & +0.03 & -6.71 & 0.40 & -0.04 & -0.01 \\\hline 
\end{tabular}
\end{center}
\end{table*}

Protonation does not only cause a charge redistribution and a variation of the energy of the frontier orbitals but it also leads to structural modifications that may impact the electronic and optical properties of the systems.
Similar effects are known to be driven also by covalently attached functional groups, as discussed in the context of graphene nanostructures~\cite{cocc+11jpcc,cocc+11jpcl,cocc+12jpcc,ivan+17jacs,hu+18jacs}.
In order to single out the contributions of such structural variations, we perform an additional set of calculations considering the neutral species in the distorted geometries obtained upon protonation. 
In these systems only the molecular backbone is considered: the proton as well as the chlorine counterion are neglected. 
In the analysis of these fictitious distorted systems, labeled by the prefix $D_a$ and $D_t$, according to the corresponding protonation site, we focus on the frontier states and on the resulting electronic gaps.

In the TAPT, where both protonation sites are available, the electronic gap reduction ascribed to distortions is -0.16~eV ($D_t$-TAPT) and -0.05~eV ($D_a$-TAPT, see Table~\ref{gaps}).
These variations have different origins in the two systems, according to the shifts of the frontier states.
In the $D_t$-TAPT, the energy of the LUMO is almost unchanged, compared to the pristine system, while the HOMO is at higher energy.
Conversely, in the $D_a$-TAPT, the energy of the HOMO is identical to the one of the pristine molecule, while the LUMO level is slightly down-shifted.

A gap reduction due to distortions occurs also in the $D_t$-TPT. 
In this case, it amounts only to -0.09~eV and results from the down-shift of the LUMO energy combined with the up-shift of the HOMO level (see Table~\ref{gaps}).
On the other hand, in the $D_a$-TAP, the gap increases by 0.03~eV compared to the pristine molecule, as a result of the lowering of both HOMO and LUMO energies.
From this analysis we can conclude that electronic hybridization and electrostatic interactions dominate over the structural distortions induced by proton uptake in determining the electronic structure of the protonated COF units.


\subsection{Optical properties}

\begin{figure}
	\centering
	\includegraphics[width=0.5\textwidth]{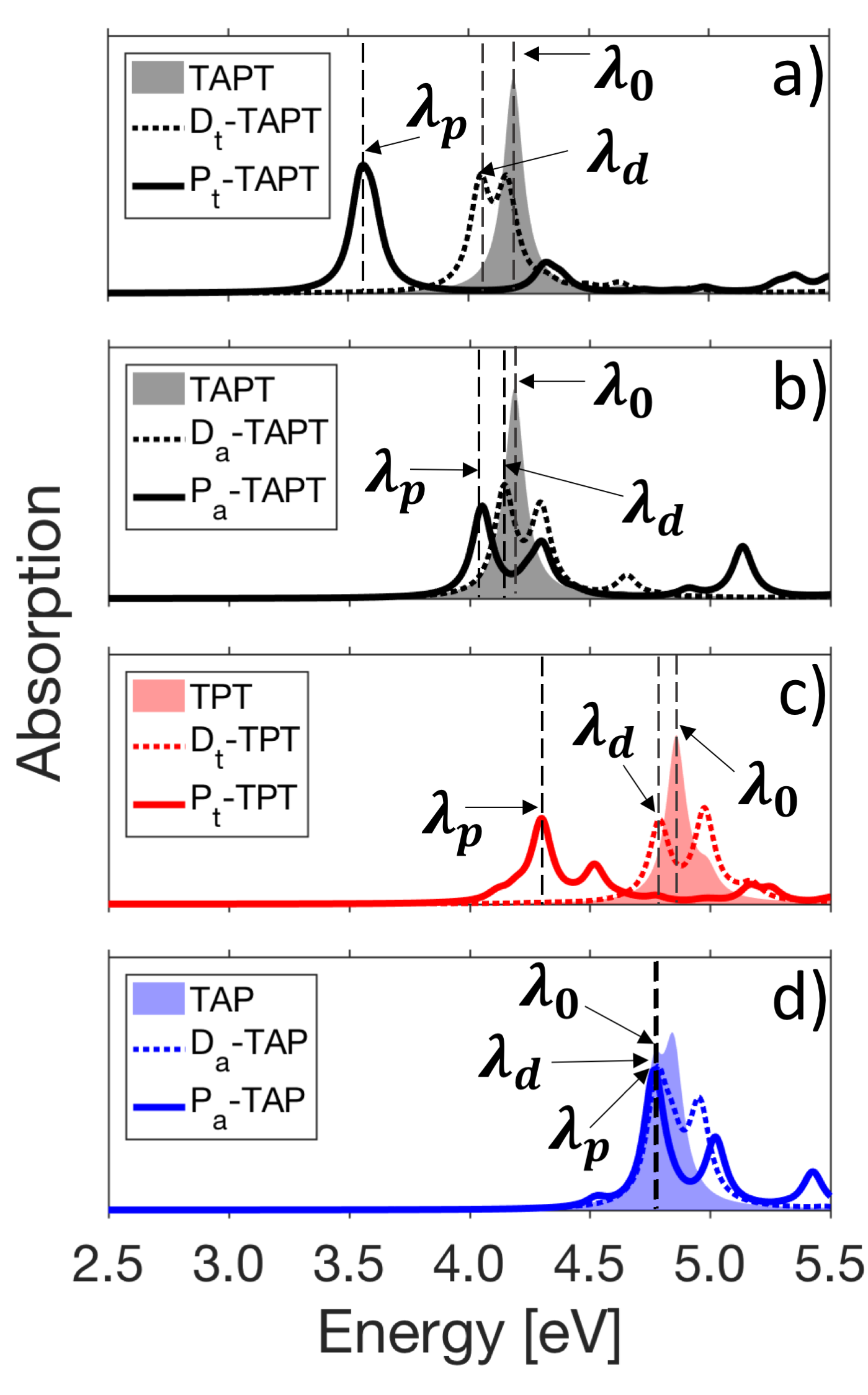}
	\caption{Absorption spectra of pristine (shaded area), protonated (solid line), and distorted (dotted lines) (a)-(b) TAPT, (c) TPT and (d) TAP in water solution. Vertical lines mark the energy of $\lambda_0$, $\lambda_p$, and $\lambda_d$. A Lorentzian broadening of 100 meV is applied to all spectra.}
	\label{abs}
\end{figure}

In the final part of our analysis, we examine the optical properties of the protonated COF units. 
The absorption spectra of the pristine molecules (Figure~\ref{abs}, shaded areas) are dominated by an intense absorption peak at the onset ($\lambda_0$), which is found at 4.19~eV in the TAPT, at 4.86~eV in the TPT, and at 4.76~eV in the TAP (see Table~\ref{composition}). 
The energies of these maxima are consistent with the trends obtained for the electronic gaps (see Table~\ref{gaps}).
In the pristine TAPT, $\lambda_0$ corresponds to two degenerate excitations, composed by the HOMO~$\rightarrow$~LUMO and the HOMO-1~$\rightarrow$~LUMO transitions (see Table~S3). 
The extended spatial overlap between these pairs of occupied and unoccupied molecular orbitals (see Figure~\ref{elestr}) explains the high oscillator strength of $\lambda_0$.
Also in the pristine TAP the onset is formed by two non-degenerate but energetically very close excitations (see Table~S3).
Finally, in the pristine TPT, the first peak is given by a double-degenerate excitation due to the symmetry of the molecule, stemming in turn from the transitions between the double-degenerate frontier states (see Figure~\ref{elestr2}).

\begin{table*}
\centering
\caption{Excitation energies ($E_{exc}$) and binding energies ($E_b$) in eV, and oscillator strength (OS) of the lowest-energy bright excitations $\lambda_0$, $\lambda_p$, $\lambda_d$ marked in Figure~\ref{abs} for the pristine, protonated, and distorted molecules, respectively. Variations of excitation and binding energies ($\Delta E_{exc}$ and $\Delta E_b$, respectively) are with respect to $\lambda_0$. }
\label{composition}%
\begin{center}
\begin{tabular}{c|cc|c|cc|c}
\hline
\textbf{Species}  & \textbf{Excitation} & E$_{exc}$ & $\Delta E_{exc}$ & E$_b$ & $\Delta E_b$ & OS \\\hline\hline
$TAPT$ & $\lambda_0^*$  & 4.19 & $-$ & 2.40 & $-$ & 2.50 \\ \hline
$P_t-TAPT$ & $\lambda_p$  & 3.55 & -0.64 & 2.20 & -0.20 & 1.29 \\
$D_t-TAPT$ & $\lambda_d$  & 4.05 & -0.14 & 2.38 & -0.02 & 1.23 \\ \hline
$P_a-TAPT$ & $\lambda_p$ & 4.05 & -0.14 & 2.18 & -0.22 & 1.11 \\
$D_a-TAPT$ & $\lambda_d$  & 4.14 & -0.05 & 2.40 & 0.00 & 1.28 \\ \hline \hline
$TPT$ & $\lambda_0^*$  & 4.86 & $-$ & 2.58 & $-$ & 1.72 \\ \hline
$P_t-TPT$ & $\lambda_p$  & 4.30 & -0.56 & 2.08 & -0.50 & 0.95 \\
$D_t-TPT$ & $\lambda_d$  & 4.79 & -0.07 & 2.56 & -0.02 & 0.82 \\ \hline \hline
$TAP$ & $\lambda_0$  & 4.76 & $-$ & 2.32 & $-$ & 0.60 \\
\hline            
$P_a-TAP$ & $\lambda_p$ & 4.76 & 0.00 & 1.93 & -0.30 & 1.14 \\
$D_a-TAP$ & $\lambda_d$  &4.78 & +0.02 & 2.33 & +0.01 & 1.00\\\hline
\end{tabular}
\end{center}
\end{table*}

Upon protonation, the optical absorption spectra are modified with respect to the neutral cases, as expected from the variations in the electronic structure discussed above. 
In all protonated species, the oscillator strength of the first peak $\lambda_p$ is reduced compared to $\lambda_0$, due to the spatial redistribution of the frontier orbitals.
On the other hand, the energy of $\lambda_p$ does not entirely reflect the trends of the electronic properties. 
While the spectra of the protonated TAPT (on both sites) and of the TPT are red-shifted in accordance to the corresponding reduction of the gaps, in the spectrum of the protonated TAP, the onset is found at the same energy as in the pristine molecule (see Table~\ref{composition}).
This result is in agreement with experiments~\cite{BojdysCOF} but it is rather counter-intuitive.
It is worth noting that $\lambda_p$ is not the lowest-energy excitation either in $P_a$-TAP or in $P_t$-TPT. A careful inspection of Figure~\ref{abs}(c)-(d) reveals the presence of a weak shoulder before the main peak in the spectra of the aforementioned protonated species. These lowest-energy excitations, which are mainly composed of HOMO $\rightarrow$ LUMO transitions (see Table~S3), have weak oscillator strength due to the reduced spatial overlap between the involved frontier orbitals (see in Figure~\ref{elestr2}b). 
This argument alone is however insufficient to explain the different spectral shifts experienced by the protonated COF units in comparison with their pristine counterparts. 

In order to reconcile the optical spectra of the protonated triazine derivatives with the results obtained for their corresponding electronic structure, we inspect the binding energy ($E_b$) of the analyzed excitations. This quantity, corresponding to the difference between the electronic gap and the excitation energy, is most accurately estimated from the solution of the BSE~\cite{rohl-loui00prb,delp+06prl,tiag-chel06prb,blas-atta11apl,fabe+12prb,brun+15jcp,cocc-drax15prb} on top of the $GW$ approximation for the electronic self-energy. However, mainly due to (partial) error compensation (see details in the SI), TDDFT results obtained with a range-separated hybrid functional can provide a reliable estimate of this quantity~\cite{refa+11prb,lee+15rscadv,sun+16jpcc,hu+17jcc,kran+17jpcc,zhu+18jpcc,li+19njc} for the purpose of this work.
In order to corroborate the reliability of our results, a direct comparison between the optical spectra computed from MBPT and hybrid TDDFT for the pristine molecules \textit{in vacuo} is reported in the Supporting Information. 
In the TAPT, the binding energy of $\lambda_p$ is about 0.20~eV lower compared to the one of $\lambda_0$, regardless of the protonation site  (see Table~\ref{composition}). 
Such a reduction is given by the combination of several factors: Proton uptake increases the electronic screening. 
Furthermore, it triggers the spatial rearrangement of the frontier orbitals that are involved in the corresponding optical transitions (see Figure~\ref{elestr}b and Table~S3).
As a result, the overlap between the photo-excited electron and hole is reduced in the protonated system compared to the pristine case.
This behavior contributes to decreasing the binding energy in addition to the oscillator strength discussed above. 
The different spectral red-shifts experienced by $\lambda_p$ in $P_t$-TAPT and in $P_a$-TAPT compared to $\lambda_0$ (-0.64~eV and -0.14~eV, respectively, see Table~\ref{composition}) can be related to the trends of the corresponding electronic gaps (see Table~\ref{gaps}).
In the TPT, when comparing $\lambda_p$ with $\lambda_0$, the variation of $E_b$ is almost as large as the variation of $E_{exc}$, which explains why the reduction of the optical gap in this protonated species is approximately half of the decrease of the electronic gap. 
Finally, in the counter-intuitive case of the TAP, our results indicate that, upon protonation, the decrease of the binding energy by -0.38~eV compensates in full the gap reduction.
Noteworthy, in this case, the variations of excitation energy and binding energy (Table~\ref{composition}) do not sum up to the variation of the electronic gap (see Table~\ref{gaps}), due to the presence of the weak excitation below $\lambda_p$ in the spectrum of the protonated species.

Finally, following the same path as in the analysis of the electronic structure, we analyze the optical behavior of the fictitious neutral species in the geometries of the protonated systems (see Figure~\ref{abs}).
In this case, the electronic and optical properties mirror each other well.
Except for the TAP, in all systems, the lowest-energy peak ($\lambda_d$) is red-shifted with respect to $\lambda_0$ in the pristine counterparts (see Table~\ref{composition}).
The amount of such a red-shift is significantly lower compared to the one experienced by $\lambda_p$, in agreement with the results obtained for the electronic gaps (see Table~\ref{gaps}).  
Interestingly, all spectra of the distorted neutral molecules exhibit a two-peak structure at the onset, due to the splitting of the electronic states at the frontier and in its proximity.
This feature, emerging also in the spectra of the protonated species, is thus driven by structural changes in the molecules.
The splitting of the first two bright excitations occurs also in the spectrum of $P_t$-TAPT, although these two features are not distinguishable in Figure~\ref{abs}a), as their energy separation of 0.05~eV is smaller than the applied broadening of 0.1~eV. 
In the TAP, the energies of $\lambda_0$ and $\lambda_p$ are the same, and $\lambda_d$ is found at 0.02~eV above them (see Table~\ref{composition}).
Based on this analysis, it is clear that distortions induced by the proton uptake do not result in any significant variations of $E_b$ (see Table~\ref{composition}), which are instead due to electrostatic interactions and electronic correlations caused by the bound H$^+$ and by the residual Cl$^-$ in its vicinity.

\subsection{Discussion and Conclusions}

The results presented in this work provide a comprehensive picture of the electronic and optical properties of protonated triazine derivatives as building blocks of COFs.
The investigation of three species, characterized by the presence of a triazine ring (TPT), amino groups (TAP), and both (TAPT), allows us to rationalize the role of the protonation site. 
Proton uptake is energetically more favorable on the triazine ring, which stabilizes the process also when it occurs on the amino group, consistent with the experimental results reported in Ref.~\citenum{BojdysCOF}.

Protonation via dissociated HCl in water solution induces a charge redistribution within the molecules, which varies upon the chemical composition of the latter. 
The concomitant presence of both triazine and amino groups in the TAPT generates an interplay between these electron-withdrawing and electron-donating units, respectively, as well as with the H$^+$ and the chlorine counterion, which leads to a non-trivial reorganization of the electron density. 
In the other considered molecules, TPT and TAP, where either the triazine ring or the amino groups are exclusively present, respectively, the extra positive charge of the proton is mainly added to the chemical group to which the H$^+$ is attached. 

In all the inspected species, protonation causes a reduction of the electronic gap.
This results from the concomitant action of two mechanisms which are most pronounced  when the proton uptake occurs on the triazine ring: (1) The reorganization of the electronic levels of the pristine species upon protonation and (2) the electrostatic attraction between the protonated species and the counterion in its proximity.
We do not anticipate that for counterions other than Cl$^-$ this general picture will be altered.

Protonated TAPT and TPT exhibit a red-shifted absorption onset compared to the pristine counterparts, while in the spectrum of protonated TAP the lowest-energy absorption maximum is almost at the same energy as in the neutral molecule.
The spectral red-shift of the absorption spectra predicted by our calculations is in qualitative agreement with the experimental results reported in Ref.~\citenum{BojdysCOF} and, more generally, with the systematic bathochromic shift observed in N-heterocyclic oligomeric/polymeric systems upon proton uptake in solution or gas-phase.\cite{Liu+14csr2019,Ascherl+7jacs2019,Zhang+5cc2016,elbe+00jpca,Xu+6njc2018,Cui+7pc2018,Welch+1jacs2011}
This behavior results from two concomitant and counteracting mechanisms,  namely  the  reduction  of  the  electronic  gap  that alone would lead to a red-shift, and the decrease of the exciton binding energy, which gives rise to a blue-shift.  These two effects counterbalance each other in a different manner in the three explored compounds:  in protonated TAPT and TPT, the reduction of the electronic gap is predominant, resulting in the red-shifted spectra compared to their pristine counterparts.
In the case of TAP, the decrease of the binding energy is exactly the same as the gap reduction upon protonation, leading to an optical gap which is unchanged with respect to the pristine counterpart. 

We have also investigated the role of structural distortions induced by the proton uptake. 
The results of this analysis indicate that such effects contribute only to a minor extent to the variation of the electronic and optical properties of protonated COF units.
Their impact is limited to a slight variation of the electronic gap on the order of hundreds of meV, and to the symmetry breaking of the TAPT and TPT molecules, which is reflected in the splitting of the main absorption peak into two maxima. 
The redistribution of the molecular orbitals upon protonation reduces the mutual spatial overlap and, consequently, diminishes the oscillator strength of the optical transitions as well as their binding energy. 
On the other hand, the structural distortions induced by proton uptake do not impact the electronic screening and, hence, the binding energies of the excitations.

The results obtained in this work can be generalized to extended covalently bonded networks of the considered molecular units. 
These structures, often formed by several layers piled up on top of each other~\cite{elka+07sci,luko+11cej,noda+19acie} are evidently stiffer than the isolated moieties considered here.
The minor role of distortions in determining both the electronic structure and the optical response upon protonation, which we found in our analysis, corroborates the robustness of our findings in this regard.
Extended COFs are also subject to decrease of spatial confinement and to enhanced screening due to the delocalized electron density therein.
For these reasons, one should expect systematically red-shifted absorption spectra and less pronounced variations of the binding energies upon protonation.
Neither effect is however expected to alter the arguments formulated in this work, which are supported by the qualitative agreement of our results with experimental studies on protonated triazine derivatives in COFs \cite{BojdysCOF,Zhang+5cc2016,Liu+14csr2019,asch+19jacs}.
 
The results and the rationale proposed in this paper provide a clear understanding of the local microscopic processes triggered by the protonation of triazine derivatives in solution and how they affect their optoelectronic properties.
As such, they are relevant in the context of COFs as well as of other conjugated networked materials obtained from solution processing.

\section*{Conflicts of interest}
There are no conflicts to declare.

\section*{Acknowledgements}
We thank Michael Bojdys for the inspiring discussions that stimulated this work.
We are grateful to Ana M. Valencia and Jannis Krumland for fruitful exchanges and suggestions and for their careful reading of this manuscript prior submission.
This work was partly funded by the Deutsche Forschungsgemeinschaft (DFG) - project number 182087777 - SFB 951. 
Financial support from Fondazione Della Riccia (M.G.) and from the Erasmus+ Programme of the European Union (E.D.A.) is kindly acknowledged. Computational resources provided by the North-German Supercomputing Alliance (HLRN), project bep00076, and by the HPC cluster CARL at the University of Oldenburg, Germany, funded by the DFG (project number INST 184/157-1 FUGG) and by the Ministry of Science and Culture of the Lower Saxony State.

\begin{suppinfo}
The analysis of the electronic and optical properties of the cationic protonated COF Units, additional information regarding the optical excitations plotted in Figure~\ref{abs} and the single-particle states contributing to them, and the study of the electronic and optical excitations of the neutral species \textit{in vacuo}, as computed from TDDFT and MBPT are reported in the Supporting Information (https://pubs.acs.org/doi/10.1021/acs.jpcc.0c08812).
\end{suppinfo}

\providecommand{\latin}[1]{#1}
\makeatletter
\providecommand{\doi}
  {\begingroup\let\do\@makeother\dospecials
  \catcode`\{=1 \catcode`\}=2 \doi@aux}
\providecommand{\doi@aux}[1]{\endgroup\texttt{#1}}
\makeatother
\providecommand*\mcitethebibliography{\thebibliography}
\csname @ifundefined\endcsname{endmcitethebibliography}
  {\let\endmcitethebibliography\endthebibliography}{}

\end{document}